\documentclass[12pt]{iopart}

\usepackage{bm}

\usepackage{amssymb}
\usepackage{graphicx}

\newcommand{\beq}{\begin{equation}}
\newcommand{\eeq}{\end{equation}}
\newcommand{\beqar}{\begin{eqnarray*}}
\newcommand{\eeqar}{\end{eqnarray*}}

\newcommand{\dt}{{\rm d}}

\newcommand{\Hh}{\hat{H}}

\newcommand{\phih}{\hat{\phi}}

\newcommand{\Pc}{\mathcal{P}}

\newcommand{\taub}{\mbox{\boldmath{$\tau$}}}

\newcommand{\rtarr}{\rightarrow}

\newcommand{\rhoh}{\hat{\rho}}

\newcommand{\nb}{{\bf n}}
\newcommand{\tb}{{\bf t}}

\newcommand{\qb}{{\bf q}}
\newcommand{\pb}{{\bf p}}

\newcommand{\rb}{{\bf r}}

\newcommand{\dg}{\dagger}

\newcommand{\lan}{\langle}
\newcommand{\ran}{\rangle}
\newcommand{\om}{\omega}

\newcommand{\De}{\Delta}
\newcommand{\la}{\lambda}

\newcommand{\eps}{\varepsilon}

\newcommand{\lt}{\left}
\newcommand{\rt}{\right}
\begin{document}

\title{Excitonic condensation in a double-layer graphene system}
\author{Maxim Yu. Kharitonov$^{1,2}$ and Konstantin B. Efetov$^2$}
\address{$^1$Materials Science Division, Argonne National Laboratory, Argonne, IL
60439, USA \\
$^2$Institut f\"ur Theoretische Physik III, Ruhr-Universit\"at Bochum, 44780
Bochum, Germany }
\date{\today}

\begin{abstract}
The possibility of excitonic condensation in a
recently proposed electrically biased double-layer graphene system
is studied theoretically.
The main emphasis
is put  on obtaining a reliable analytical estimate for the transition temperature into the excitonic state.
As in a double-layer graphene system the total number of fermionic
\textquotedblleft flavors" is equal to $N=8$ due to two projections of spin,
two valleys, and two layers, the large-$N$ approximation appears to be  especially
suitable for  theoretical investigation of the system.
On the other hand,
the large number of flavors makes
screening of the bare Coulomb interactions very efficient,
which, together with the suppression of backscattering in graphene, leads to an extremely low energy
of  the excitonic condensation.
It is shown that the effect of screening on the excitonic pairing is just as
strong in the excitonic state as it is in the normal state.
As a result, the value of the excitonic gap $\De$ is found to be in full
agreement with the previously obtained estimate for the mean-field transition temperature $T_c$,
the maximum possible value $\De^{\rm max},T_c^{\rm max}\sim 10^{-7} \epsilon_F$ ($\epsilon_F$ is the Fermi energy) of both being
in $  1{\rm mK}$ range for a perfectly clean system.
This proves that the energy scale  $\sim 10^{-7} \epsilon_F$
really sets the upper bound for the transition temperature and invalidates the recently expressed conjecture about the high-temperature
first-order transition into the excitonic state. These findings suggest
that, unfortunately, the excitonic condensation in graphene double-layers
can hardly be realized experimentally.
\end{abstract}

\maketitle

\section{Introduction}

The idea of excitonic condensation in metallic systems was
originally proposed~\cite{KK} by Keldysh and Kopaev for semimetals with
overlapping conduction and valence bands. They have shown that the
attractive Coulomb interaction between electrons and holes leads to an instability towards formation of bound electron-hole
pairs, analogous to the Cooper instability in superconductors.
Somewhat later, Lozovik and Yudson suggested\cite{LY}
that excitonic condensation could be realized in a double-layer system of spatially separated
electrons and holes. If the layers are close enough to each other, the
interlayer Coulomb interaction could still be
appreciable, which would lead to coupling between electrons and holes.

It took a while after the proposals of Refs.~\cite{LY}
until the technology able to fabricate electron-hole
bilayers has been developed. So far, experimental efforts
have been mainly concentrated on the
GaAs/Al$_{x}$Ga$_{1-x}$As
double quantum well heterostructures
and the evidence~\cite{exp1,exp11,exp12,exp2,exp3,exp4,exp5,exp6} for excitonic condensation in such systems
based on the investigation of the Coulomb drag~\cite{exp1,exp11,exp12} and photoluminescence
spectra~\cite{exp2,exp3,exp4,exp5,exp6} is building.
Another exciting phenomenon in the coupled semiconductor bilayers is the quantum Hall ferromagnetism~\cite{Eisen,EMD}.

\begin{figure}
\includegraphics[width=.70\textwidth]{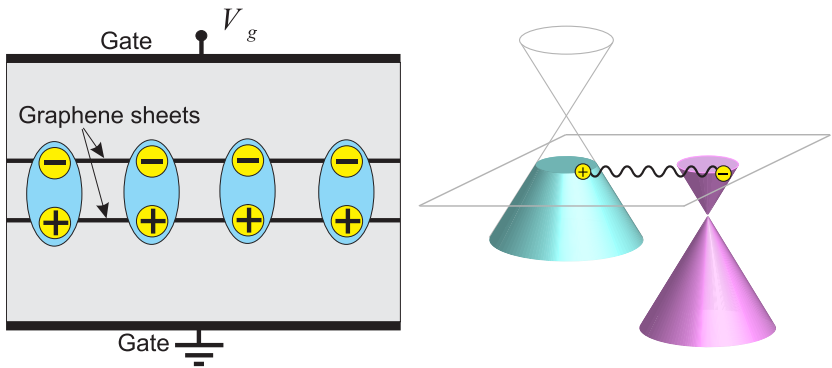}
\caption{ Excitonic condensate in a system of two spatially separated
graphene layers. Electrons and holes in the layers are induced by applying
the external gate voltage and the excitonic pairing is caused by the attractive interlayer Coulomb interaction.}
\label{fig:system}
\end{figure}

Besides semiconductor double quantum wells, what other materials and technology could be used
for the search of the excitonic condensation?
Graphene~\cite{gr1,gr2,gr3,gr4,gr5,gr6,grrise}, an isolated atomic monolayer of carbon,
could seem very attractive for this purpose.
Indeed, quite recently several groups of authors~\cite{LS,LMS,MD,ZJ} proposed
the following graphene-based setup (Fig.~\ref{fig:system}) as a candidate
for the observation of the excitonic condensation.
If one takes two graphene layers separated by
an insulator, then electron doping in one layer and hole doping in the other can be
obtained by applying external gate voltage.
Analogously to the semiconductor quantum double wells,
the interlayer Coulomb interaction would lead to coupling between
electrons and holes.

What attractive properties does such graphene system have?
First, the relatively high values of the Fermi energy $\epsilon _{F}\sim 0.3\,%
{\rm eV}$ that can be achieved in graphene by using electric gates~\cite{gr2} is an obvious advantage, since
the condensation temperature is expected scale linearly with $\epsilon _{F}$. %
Second, as any mismatch between the Fermi momenta of electrons and holes tends to
suppress the condensate in the same way as the Zeeman
splitting suppresses the conventional superconductivity, the nearly
perfect matching of the electron and hole branches of the spectrum and the
ability to fine-tune the carrier density become extremely important.

So, in terms of engineering properties, the double-layer graphene
system~(Fig.~\ref{fig:system}) could seem exceptionally
attractive
for realization of the excitonic condensation. %
However, before one proceeds with the fabrication of this system,
an important theoretical question that should be answered is
how high the  temperature of transition to the excitonic state
in such  a setup could be.

Some seriously conflicting predictions have been reported in the literature
on this matter.  Initially, an estimate for the Berezinski-Kostelitz-Thouless (BKT) transition~\cite{Bereza,KT} temperature
in a double-layer graphene system was made in
Refs.~\cite{MD,ZJ}. Studying the problem numerically,  the authors came to conclusion that the
BKT temperature $T_{KT}^{\ast }$ in
this system could be very
large,
\begin{equation}
T_{KT}^{\ast }\sim 0.1\epsilon _{F}.  \label{eq:TBKT*}
\end{equation}%
With the Fermi energy $\epsilon _{F}\sim 0.3{\rm eV}$, this would
correspond to room temperatures, $T_{KT}^{\ast }\sim 300{\rm K}$,
which would be very encouraging for experimentalists.

Unfortunately, the estimate (\ref{eq:TBKT*}) for the transition temperature  is too optimistic. As pointed out in Ref.~\cite{KE} this result
stemmed from the complete neglect of
screening of the interlayer Coulomb interaction by the carriers
in the layers in the analysis of Refs.~\cite{MD,ZJ}. As the bare Coulomb interactions in graphene are not
weak, it is not surprising that neglecting screening the transition temperature appeared to be  only one order smaller than
the Fermi energy $\epsilon _{F}$. In Ref.~\cite{KE} the mean field
transition temperature $T_{c}$, at which the normal state becomes unstable towards electron-hole pair
formation, was calculated.
It was demonstrated that when screening is taken into account the highest possible value of the mean-field
temperature turns out to be extremely small,
\beq
    T_c^{\rm max} \sim 10^{-7} \epsilon_F.
\label{eq:Tcmax}
\eeq
The reason for that is the large number $N=8$ of ``flavors'' in a double-layer graphene system, which makes screening very efficient, and
the suppression of backscattering in graphene due to the chiral nature of Dirac quasiparticles.
These factors combined make the system effectively weakly interacting, and the transition temperature depends
exponentially on the coupling constant in the weak-coupling regime.
Since the thermal fluctuations of the order parameter in two
dimensions can only suppress the superfluid properties of the system, the
BKT transition temperature $T_{KT}$ can be only smaller than the mean-field temperature (\ref{eq:Tcmax}).

Somewhat later, in Ref.~\cite{MDc}, the authors of Ref.~\cite{MD} presented
arguments to justify the neglect of screening in Refs.~\cite%
{MD,ZJ}. They argued that in the excitonic state screening had to be essentially suppressed due to the gap in the excitation spectrum.
This would mean that
in the excitonic phase, at least at low enough temperatures, the interactions are not screened and  the system
is in the strong coupling regime of the  bare Coulomb interactions.
Therefore, the value of the gap  $\De^*(T=0)$
at zero temperature, when the suppression of screening is maximal,
had to be of the same order as the initial estimate (\ref{eq:TBKT*})
for the BKT temperature,
\beq
    \De^*(T=0) \sim 0.1 \epsilon_F.
\label{eq:De*}
\eeq
A strong discrepancy between the small value (\ref{eq:Tcmax}) of the mean-field
temperature and the large value (\ref{eq:De*}) of the gap in the excitonic state implies, according to the authors of Ref.~\cite{MDc}, that the
system should undergo a first order transition at temperature $T \sim \De^*(T=0)$, which would again
correspond to room temperatures.

The purpose of this paper is to review in more detail the arguments
leading to Eqs.~(\ref{eq:Tcmax}) and (\ref{eq:De*}) in order to
understand which of the estimates for the transition temperature is finally correct.
 We will show that, in terms of its effect on the
electron-hole pairing, screening in the excitonic state is just as strong as it is in the normal state.
This is because a wide range of scattering momenta of electron and hole in a
pair 
contributes to the value of the excitonic gap,
whereas screening is
suppressed only in a much narrower range of small momenta. As a result,
an accurate analysis of the excitonic state
yields essentially the same estimate
\begin{equation}
\De^{\rm max}(T=0)\sim 10^{-7}\epsilon _{F}  \label{eq:Demax}
\end{equation}%
for the highest possible value of the gap
as the one obtained for the mean field temperature~[Eq.~(\ref%
{eq:Tcmax})].
Therefore, the energy scale $\sim 10^{-7}\epsilon _{F}$ is the only one
in the problem, no matter whether one considers the normal or
excitonic state, and Eq.~(\ref{eq:Tcmax}) is indeed the upper bound
for the transition temperature into the excitonic state.
This invalidates the argument of
Ref.~\cite{MDc} about the high-temperature first-order transition.
Considering that the excitonic condensate is sensitive to the impurity scattering,
such a low transition temperature renders the observation of the excitonic condensation
 in a double-layer graphene system very improbable.

The analysis of the problem
presented in this paper is largely based on Ref.~\cite{AKT}, in which a detailed theory of excitonic condensation in a single-layer graphene subject to the parallel magnetic field was developed.
Although here we mainly concentrate on obtaining an analytical estimate for the transition temperature,
many of the fine features of the excitonic state studied in Ref.~\cite{AKT} could be carried over to the double-layer graphene system.

\section{Large-$N$ approach to a double-layer graphene system}

In this section, we will
outline the main idea of the theoretical method, the large-$N$
approximation, which will be used to obtain analytical estimates (\ref%
{eq:Tcmax}) and (\ref{eq:Demax}) for the transition temperature and the excitonic gap.
For a double layer graphene system the large-$N$ approximation appears to be
particularly reliable and is expected to provide good quantitative
predictions.

\begin{figure}
\includegraphics[width=.70\textwidth]{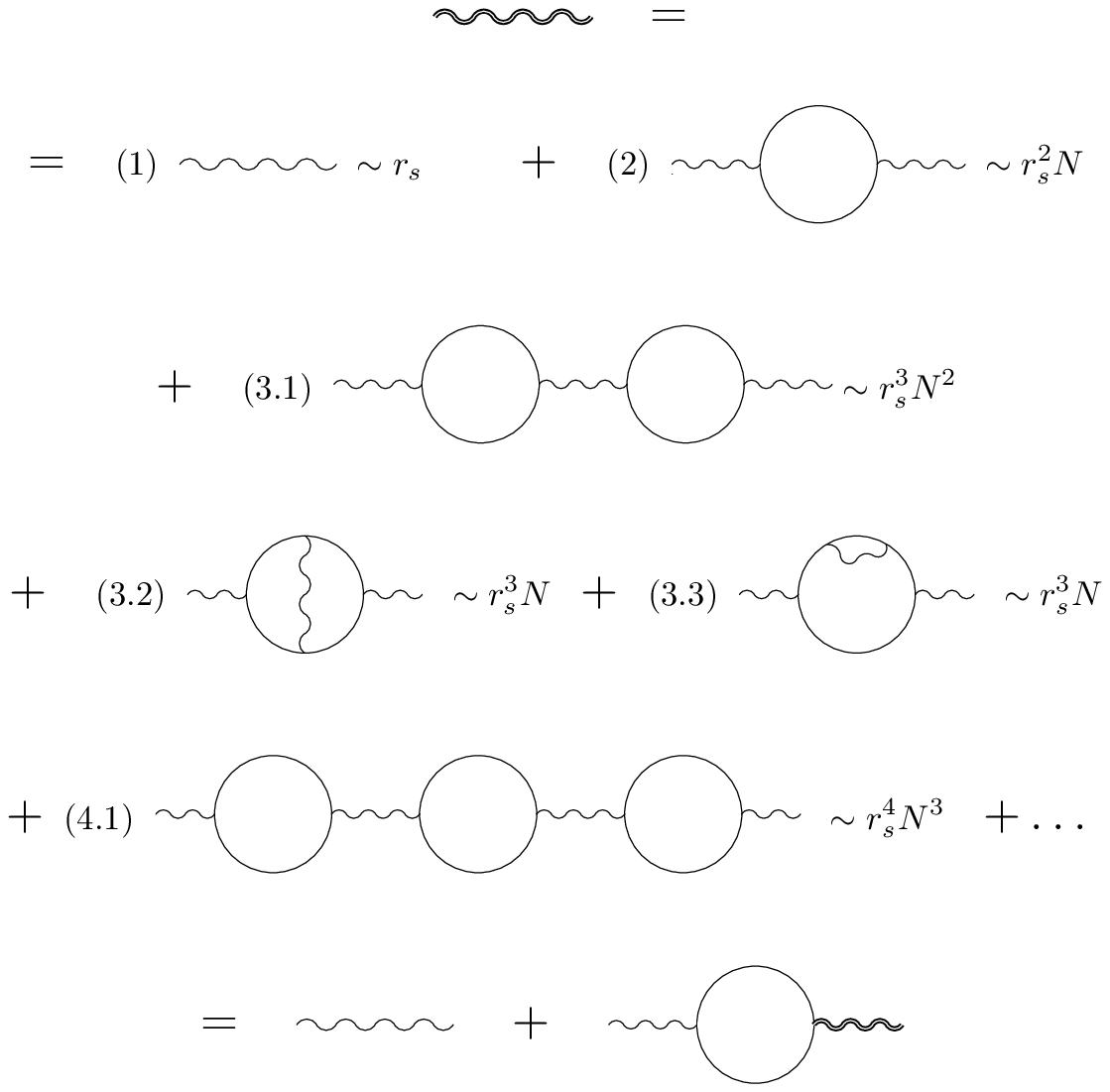}
\caption{The idea of the large-$N$ approximation.
In each order in the bare Coulomb interaction (wavy lines),
the leading in $N\gg 1$ diagrams for the effective interaction (``photon propagator'', double wavy line)
are those that contain the maximum number of fermionic loops.
So, in the third order [diagrams (3.1)-(3.3)], the diagrams (3.2) and (3.3) are smaller than (3.1) in $1/N \ll 1$ and can be neglected.
Performing the same analysis for higher orders, one arrives at the series analogous to that of the RPA.
The summation of this series can be reduced to solving an equation for the effective interaction (last line). }
\label{fig:largeN}
\end{figure}

The bare strength $r_{s}=e^{2}/(\varepsilon v)$ ($e$ is the electron charge, $\varepsilon $ is the
dielectric constant of the insulating medium surrounding graphene, $%
v\approx 10^{8}{\rm cm/s}$ is the velocity of the Dirac spectrum, and we will use the units, in which the Planck's constant  $\hbar=1$ throughout the paper) of
the Coulomb interactions in graphene is not that small.
For graphene in vacuum ($\varepsilon =1$) $r_{s} \approx 2.16 $  and
for SiO$_{2}$ as an insulator
typically
$r_{s}\approx 1$ or somewhat smaller.
These values of $r_{s}$ are not exceptionally large,
but cannot be considered as small either.
Therefore, strictly speaking, the weak-coupling
treatment of the Dirac fermions in graphene for the realistic values of $%
r_{s}$ 
is not applicable.
In the diagrammatic approach, for $r_{s}\gtrsim 1$ one would have to sum up
all diagrams in each order in the bare interaction. Clearly, this task  cannot be
carried out analytically. 
If, however, the number $N$ of independent fermionic species, or \textquotedblleft
flavors\textquotedblright\, is large, one has an additional expansion
parameter, which can efficiently be exploited.
The flavors can correspond to, e.g., different projections of the real spin $S$ of particles, in which case
$N$ would be the total number of possible spin projections,
$N=2S+1$, or to some other effective discrete degrees of freedom, which stem
from the properties of the particles spectrum.

So, if the number of flavors $N$ is indeed large, one can notice (Fig.~\ref{fig:largeN}) that after summing the contributions of
all flavors (``performing the trace over the flavor indices'') each closed fermionic loop
consisting of the fermionic Green's functions gives the factor $N$.
Therefore, in a given order in the bare interaction,
the leading in $N\gg 1$ contribution to the effective interaction (``photon propagator'') comes from
the diagrams with the maximum numbers of electron loops. These are the
diagrams that contain only the polarization bubbles consisting of two
Green's functions. The series of such diagrams can already
be summed up as it is \emph{formally} identical to the one of the well-known Random Phase Approximation (RPA).

Physically, the leading in $N$ series describes the effect of linear
screening of  interactions by the carriers. %
Of course, screening is to some extent present in any
system of free charged particles. However, in the limit of large $N$ screening is particularly strong,
since \emph{all} $N$ species participate in the screening of the
interactions between fermions of \emph{each} particular species. Screening  makes
the system effectively weakly interacting reducing the coupling constant from $r_s$ to $1/N$: $r_s \rtarr 1/N$.
This allows one to study the system
analytically, despite the fact that the bare Coulomb interactions may be not
weak ($r_s \gtrsim 1$).

How large is the number of flavors $N$ in a
double-layer graphene system (Fig.~\ref{fig:system})? In a single layer of
graphene the number of species is equal to
\[
N_{1}=N_{s}N_{v}=4
\] due to $N_{s}=2$
possible projections of spin and $N_{v}=2$ valleys.
The large-$N$ approach to a single-layer graphene was used in a number of works~\cite{AKT,YS,Son,FA,MFS,Khvesh,GGMS,Khvesh2,Khvesh3} before.
For a double-layer system one has an additional \textquotedblleft
which-layer\textquotedblright\ degree of freedom, since each carrier
can belong to either one of the layers. This twofold ($N_{l}=2$) degeneracy
appears formally in calculations when summing the polarization
bubbles of both layers.
Therefore, in a double-layer graphene system (Fig.~\ref{fig:system}) the
total number of flavors is equal to
\begin{equation}
N=N_{s}N_{v}N_{l}=2^{3}=8.  \label{eq:N}
\end{equation}%
For this value it is quite reasonable to expect the large-$N$ approximation to
give good quantitative predictions. The accuracy of the obtained results
will be estimated below in Sec.~\ref{sec:mf}.

So, with $N=8$ the large-$N$ approximation seems to be particularly suitable
for the double-layer graphene system. On the one hand, this is an
attractive situation for theorists, since one is able to perform
analytical calculations and make reliable quantitative predictions for a
system of non-weakly interacting electrons. On the other hand, since the system is
effectively in the weak coupling regime, one could expect the transition
temperature to be quite small, which  certainly is a pessimistic prospective
for experimentalists trying to observe excitonic condensation in such
a system.

Before we proceed with calculations, we would like
to emphasize one important point. 
Although the diagrammatic series in RPA and large-$N$ approximation are
formally identical, there is an important difference in the validity of the
two approaches. The RPA works, when the characteristic energies $\epsilon $
and momenta $q$ (set, e.g., by temperature) of the excitations contributing
to the studied quantity are small compared to the Fermi scale: $\epsilon \ll
\epsilon _{F}$ and $q\ll p_{F}$. The reason for that is the possibility
to describe the low-energy excitations in terms of the charge and spin
densities, quadratic in the fermionic operators.
One the other hand, the large-$N$ approximation is
applicable as long as $N\gg 1$ and does not require the smallness of the
relevant energy scales compared to the Fermi energy. It works even when the
Fermi surface is absent in the problem, e.g., in the vicinity of the Dirac
point in graphene\cite{AKT,YS,Son,FA,MFS,Khvesh,GGMS,Khvesh2,Khvesh3}.
Actually, as we will see in the next section, the excitonic gap and the
transition temperature 
are determined by all transfer momenta $q$ in the range $0\leq q\leq 2p_{F}$, where $p_F$ is the Fermi momentum.
Therefore, strictly speaking, the RPA does not apply to the problem when the bare
interactions are not weak. %
It does, however, apply to the case of weak bare
interactions and we will come back to this point, when we discuss the
transition temperature in more detail in Sec.~\ref{sec:mf}.

\section{Gap equation \label{sec:gap}}
In this section, we present the self-consistency equation for the excitonic
gap, which will be used in the next two sections to obtain the mean-field temperature and the gap
in the excitonic state.

To study the excitonic condensation in a double-layer graphene system~(Fig.~\ref{fig:system}) we will use the following electron Hamiltonian,
\begin{equation}
\Hh=\sum_{i=e,h}\Hh_0^i+\sum_{i,j=e,h}\Hh_{\rm int}^{
ij }.  \label{eq:H}
\end{equation}%
Here, the indices $i$ and $j$ numerate  electron ($e$) and hole ($h$) layers.
The terms
\beq
\Hh^i_0= \int \dt \rb \, \phih^\dg_i (\rb ) ( v  \taub \pb-\mu_i) \phih_i(\rb)
\label{eq:H0}
\eeq
describe the free Dirac particles in  graphene layers, where $\phih_i(\rb)$ are the Dirac spinor field operators,
$\rb$ is a two-dimensional radius vector in graphene layers,
$\pb=-{\rm i} \nabla_\rb$ is the momentum operator, and $\taub=(\tau_x,\tau_y)$ is a vector consisting of the Pauli matrices $\tau_x$ and $\tau_y$ in the sublattice space of graphene honeycomb lattice.
The chemical potentials $\mu_{e,h}$ of electrons and holes have opposite signs and will be assumed
equal in absolute value,
\[
    \mu_e=-\mu_h=\epsilon_F.
\] This corresponds to the most favorable
situation for the excitonic condensation, as any mismatch $\mu_e - |\mu_h|$  acts as an effective Zeeman field
and tends to suppress the condensate. Further, the terms
\beq
    \Hh_{\rm int}^{ij}=\frac{1}{2}\int \dt \rb \, \dt \rb'\, \phih^\dg_i(\rb) \phih^\dg_j(\rb') V_{ij}(\rb-\rb')\phih_j(\rb')\phih_i(\rb)
\eeq
in Eq.~(\ref{eq:H}) describe the intralayer ($i=j$)  and interlayer ($i \neq j$) Coulomb interactions between the Dirac particles
with $V_0(\rb)=V_{ee}(\rb)=V_{hh}(\rb)$ and $V_1(\rb)=V_{eh}(\rb)=V_{he}(\rb)$
given by the Coulomb potential
\beq
    V_0(\rb)=\frac{e_*^2}{|\rb|}, \mbox{ } V_1(\rb)=\frac{e_*^2}{\sqrt{|\rb|^2+d^2}},
\label{eq:Vcr}
\eeq
In Eq. (\ref{eq:Vcr}), $e_{\ast }$ is an
effective electron charge screened by the insulator embedding graphene
sheets, $e_{\ast }^{2}=e^{2}/\varepsilon ,$ and $d$ is the
distance between the layers.

The order parameter of the excitonic condensate is defined as
\beq
    \hat{\De}(\nb)=\int \dt \rb \, e^{-{\rm i} p_F \nb (\rb-\rb')} V_1(\rb-\rb') \lan \phih_e(\rb) \phih_h^\dg(\rb') \ran.
\eeq
Here, $\nb$ is the unit vector representing the direction of the electron momentum on the Fermi surface.
In the sublattice space $\hat{\De}(\nb)$ has the following matrix structure,
\beq
    \hat{\De}(\nb)=\De\, \bar{\Pc}(\nb)+\De'\,  \bar{\Pc}(-\nb),
\label{eq:Deh}
\eeq
where $\bar{\Pc}(\nb)=({\rm i} \tau_z+ \taub \tb)/2$ and  $\tb=(-n_y,n_x)$ is the vector orthogonal to $\nb=(n_x,n_y)$ in the $xy$ plane.
The constant $\De$ is the excitonic gap of the spectrum, $\epsilon_p=\sqrt{\De^2+\xi^2}$, $\xi=v(p-p_F)$.
The physical quantities of the condensate are determined by $\De$, whereas the other constant in Eq.~(\ref{eq:Deh}),
$\De'$, which is related to $\De$, does not enter any of them.
The properties of the excitonic gap $\De$ will be of our main interest from now on.

Within the large-$N$ approach, the theory
is analogous to the conventional weak-coupling Bardeen-Cooper-Schriffer (BCS) theory with the
interaction given by the screened interlayer Coulomb potential. The equation for the
excitonic gap $\Delta $ has the conventional BCS-like
form~\cite{AKT}
\begin{equation}
\Delta =\lambda _{\Delta }T\sum_{n=-\infty}^{+\infty}\int_{-\epsilon _{F}}^{+\epsilon
_{F}}\dt\xi \frac{\Delta }{\varepsilon _{n}^{2}+\xi ^{2}+\Delta ^{2}}.
\label{eq:Deeq}
\end{equation}%
In Eq.~(\ref{eq:Deeq}), $T$ is the temperature, $\varepsilon _{n}=\pi
T(2n+1)$ are the fermionic Matsubara frequencies and the summation
should be carried out over all integer $n$.
The integration over
$\xi $, which diverges logarithmically if the summation over $n$ is performed first,
should be cut  by the Fermi energy $\pm \epsilon _{F}$.
Further, the dimensionless
coupling constant of the screened interlayer Coulomb interactions is given by the expression
\begin{equation}
\lambda _{\Delta }=\nu \int_{0}^{2\pi }\frac{\dt\theta }{2\pi }V_{\Delta
}\left( 2p_{F}\sin \frac{\theta }{2}\right) \frac{1+\cos \theta }{2},
\label{eq:la}
\end{equation}%
where $\nu =\epsilon _{F}/(2\pi v^{2})$ is the density of states in each
graphene layer per one projection of spin and one valley.
In Eq.~(\ref{eq:la}), $V_{\Delta
}(q)$ is the Fourier component of the static (at frequency $\om=0$) screened interlayer Coulomb interaction potential.
The static limit for the coupling potential is valid with the logarithmic accuracy.
The potential $V_\De(q)$
depends on the value of the gap $\Delta $, since screening is affected by
the presence of the excitonic gap. We will discuss the latter point in detail and
present the explicit expression for $V_{\Delta }(q)$ in the excitonic phase
in Sec.~\ref{sec:BKT}.

Two important points should be emphasized about Eqs.~(\ref%
{eq:Deeq}) and (\ref{eq:la}). First, as follows from Eq.~(\ref{eq:la}), the
coupling strength $\lambda _{\Delta }$ is determined by all transfer momenta
$q=2p_{F}\sin (\theta /2)$ in the range $0\leq q\leq 2p_{F}$, or
equivalently, by all scattering angles in the range $0\leq \theta \leq 2\pi $%
. This integration over the scattering angles is standard and it also
appears in the conventional BCS theory. %
This fact is, however, crucial for obtaining the correct estimate for the
gap $\Delta $ in the excitonic state.
It will be key to disproving the statement made in Ref.~\cite{MDc}
that, in terms of its effect on the magnitude of the gap, screening is suppressed in the  excitonic state.

The second point is the presence of the factor $(1+\cos \theta )/2$ in Eq.~(%
\ref{eq:la}). It stems from the chiral nature of Dirac fermions in graphene
and leads to the suppression of backscattering. This feature is
specific to graphene and is absent in the system with
conventional electronic spectrum. So, for any point-like interaction, $%
V_{\Delta }(q)=U_{0}={\rm const}$, the coupling strength $\lambda _{\Delta
}=\nu U_{0}/2$ is actually two times smaller than the
conventionally defined dimensionless coupling constant $\nu U_{0}$. As the
transition temperature following from Eq.~(\ref{eq:Deeq}) depends on $%
\lambda _{\Delta }$ exponentially (see next section), this reduction by the
factor $2$ leads to an additional substantial suppression of the
transition temperature compared to the double layer system with
the conventional electronic spectrum.

Equations (\ref{eq:Deeq}) and (\ref{eq:la}) supplemented by the
proper form of the interlayer Coulomb potential $V_{\Delta }(q)$ allow one
to obtain an analytical estimate for the transition temperature and the excitonic gap.
In the next section we will calculate the mean field transition temperature.
The results of the next section reproduce those of Ref.~\cite{KE}. %

\section{Mean-field transition temperature \label{sec:mf}}

In order to find the mean field transition temperature $T_{c}$, one needs to linearize the
gap equation (\ref{eq:Deeq}) in $\Delta $ and determine the
temperature at which a nonzero solution to it appears. Calculating the
integral over $\xi $ and the sum over $n$ in the linearized Eq. (\ref{eq:Deeq}), one arrives at the
BCS-like expression
\begin{equation}
T_{c}\approx \exp (-1/\lambda _{\Delta =0})\epsilon _{F}.  \label{eq:Tc}
\end{equation}%
In Eq.~(\ref{eq:Tc}), the screened Coulomb potential $V_{\Delta =0}(q)$ [see Eq.~(\ref{eq:la})
for $\lambda _{\Delta =0}$] should be determined in the normal state, when $\Delta =0$.

Calculating the potential $V_{\Delta =0}(q)$ in the large-$N$ approximation is formally analogous to the RPA.
One important difference, however, is that since the relevant transfer momenta $q=2p_{F}\sin \theta /2$ belong to the range $%
0\leq q\leq 2p_{F}$ [Eq.~(\ref{eq:la})], one has to use the exact expression
for the polarization operator $\Pi _{\Delta }\left( \omega ,q\right) $,
and not its approximate form at $q\ll p_{F}$. 
However, the static polarization operator $\Pi _{\Delta =0}(q)=\Pi _{\Delta
=0}(\omega =0,q)$ in the normal state in graphene~\cite{GGMS,PO} does not
depend on momentum $q$ at all in this range 
and equals
\begin{equation}
\Pi _{\Delta =0}(q)=N_{s}N_{v}\nu ,\mbox{ }q\leq 2p_{F}.  \label{eq:Pi0}
\end{equation}%
The expressions for the bare intralayer and interlayer Coulomb interaction
potentials ~[Eq.(\ref{eq:Vcr})] in the Fourier representation read
\begin{equation}
V_{0}(q)=\frac{2\pi e_{\ast }^{2}}{q},\mbox{ }V_{1}(q)=\frac{2\pi e_{\ast
}^{2}\exp (-qd)}{q},  \label{eq:Vc}
\end{equation}
respectively. Using Eqs.~(\ref{eq:Pi0}) and (\ref{eq:Vc})
and performing analogous to the RPA calculations one obtains for the
screened interlayer Coulomb potential in the normal state
\begin{equation}
V_{\Delta =0}(q)=\frac{2\pi e_{\ast }^{2}\exp (-qd)}{q+2\varkappa +\varkappa
^{2}[1-\exp (-2qd)]/q},\mbox{ }q\leq 2p_{F}.  \label{eq:V0}
\end{equation}%
In Eq.~(\ref{eq:V0}),
\begin{equation}
\varkappa =2\pi N_{s}N_{v}e_{\ast }^{2}\nu  \label{eq:ka}
\end{equation}%
is the Debye screening wavevector (inverse screening length) in each layer.

Substituting 
Eq.~(\ref{eq:V0}) into Eq.~(\ref{eq:la}), one obtains the mean field
transition temperature $T_c$ from Eq.~(\ref{eq:Tc}).
Let us estimate the highest possible value of $T_c$.
According to Eq.~(\ref{eq:V0}) 
the potential $V_{\Delta=0 }(q)$ is
maximal, 
when the interlayer spacing $d$ 
satisfies the condition
\begin{equation}
\varkappa d\ll 1.  \label{eq:cond}
\end{equation}%
The other required condition, $p_{F}d\ll 1$, is automatically satisfied for $%
r_{s}\gtrsim 1$ in the limit (\ref{eq:cond}), since $\varkappa =r_{s}N_{s}N_{v}p_{F}=4r_{s}p_{F}$.
In reality, the limit (\ref{eq:cond}) is hardly achievable. Indeed,
on the one hand, one would like to have a large Fermi energy $%
\epsilon _{F}$ to have a higher cutoff energy in Eq.~(\ref%
{eq:Tc}), but on the other hand, higher $\epsilon _{F}$
would require very smaller interlayer distance $d$. For $\epsilon _{F}=0.3%
{\rm eV}$ and $r_{s}=1$, the condition (\ref{eq:cond}) corresponds to $d\ll
0.5\,{\rm nm}$.

So, in the limit (\ref{eq:cond}) of ``zero'' interlayer distance  $d$, the interlayer potential (\ref%
{eq:V0}) is maximal  and equals
\begin{equation}
V_{\Delta =0}^{{\rm max}}(q)=\frac{2\pi e_{\ast }^{2}}{q+2\varkappa }.
\label{eq:V0max}
\end{equation}%
Equation~(\ref{eq:V0max}) has the same form as the screened
Coulomb potential of a single layer. Note, however, the factor $2$ in the
term $2\varkappa $ in Eq.~(\ref{eq:V0max}). It originates from summing
the polarization operators of both layers and reflects the
above discussed \textquotedblleft which-layer\textquotedblright\ degeneracy~(%
$N_{l}=2$), while the Debye wavevector $\varkappa $~[Eq.~(\ref{eq:ka})]
contains the spin ($N_{s}=2$) and valley ($N_{v}=2$) degeneracies of each layer. One may say that
screening in a double-layer system is twice as efficient as in a
single layer, since the carriers of both layers participate in it.
The potential $V_{\Delta =0}^{{\rm max}}(q)$ is a decreasing function of $q$
with the maximum at $q=0$. Using Eq.~(\ref{eq:ka})
we obtain
\begin{equation}
\nu V_{\Delta =0}^{{\rm max}}(q=0)=\frac{1}{N}=\frac{1}{8}.
\label{eq:V0maxq0}
\end{equation}%
We see that the maximum (\ref{eq:V0maxq0}) of the
dimensionless coupling constant is given by the inverse number of flavors
and is, therefore, small for large $N$.
Inserting Eq.~(\ref{eq:V0maxq0}) into Eq.~(\ref{eq:la}), one obtains the
maximum possible value of the coupling constant $\lambda _{\Delta =0}$ in
the normal state: %
\begin{equation}
\lambda _{\Delta =0}^{{\rm max}}=\frac{1}{2N}=\frac{1}{16}.
\label{eq:lamax}
\end{equation}%
As discussed above, the coupling constant $\lambda _{\Delta =0}^{%
{\rm max}}$, which determines the transition temperature, is
twice as small as the one given by Eq.~(\ref{eq:V0maxq0}) due to
the suppression of backscattering. Inserting Eq.~(\ref%
{eq:lamax}) into Eq.~(\ref{eq:Tc}), one obtains the maximum possible value
of the mean-field transition temperature,
\begin{equation}
T_{c}^{{\rm max}}\approx \exp (-2N)\epsilon _{F}=\exp (-16)\epsilon
_{F}\approx 10^{-7}\epsilon _{F}.  \label{eq:Tcmax2}
\end{equation}%
We see that, although the value $\lambda _{\Delta =0}^{{\rm max}}$ [Eq. (%
\ref{eq:lamax})] is not particularly small, the mean-field temperature
appears to be extremely small due to its exponential dependence on $\lambda
_{\Delta =0}^{{\rm max}}$. With $\epsilon _{F}\sim 0.3{\rm eV}$, Eq.~(\ref%
{eq:Tcmax2}) corresponds to $T_{c}^{{\rm max}}\sim 1{\rm mK}$.

Let us here emphasize the difference between the large-$N$ approximation and the
RPA in connection to the considered problem. Basically, Eqs.~(\ref{eq:Deeq}) and (\ref{eq:la}) of the
weak-coupling BCS-like theory are valid whenever the dimensionless coupling constant $%
\lambda _{\Delta }$ of the screened interlayer Coulomb interactions is
small. In case of a sufficiently large number $N$ of flavors
the coupling constant $\lambda _{\Delta =0}^{{\rm max}}$ [Eq.~(\ref%
{eq:lamax})] is small in $1/N \ll 1$ and the applicability of the weak-coupling theory
[Eqs.~(\ref{eq:Deeq}) and (\ref{eq:la})] is justified for arbitrary strength
 of the bare  interactions. However, if the number of flavors were not
large (e.g., if $N_s=N_v=1$), Eqs.~(\ref{eq:Deeq}) and (\ref{eq:la}) would still be valid in the
weak coupling regime of the bare interlayer interaction $V_1(q)$, i.e., when
$r_{s}\ll 1$ or $p_{F}d\ll 1$, see  Eq.~(\ref{eq:Vc}).
Indeed, in the integral over $q=2p_{F}\sin (\theta /2)$ in Eq.~(\ref{eq:la}%
), $\nu V_{1}(q\sim p_{F}) \sim r_{s}\ll 1$ in the major range
of integration $q\sim p_{F}$, and for these $q$ the bare interaction
potential is weak. Only for $q\ll p_{F}$ the bare coupling $\nu V_{1}(q\ll
p_{F})\gtrsim 1$ is not small. But for small $q\ll
p_{F}$ the RPA is valid and the effective interaction is described by Eq.~(%
\ref{eq:V0}), in which the number of flavors $N$ does not need to be large.
Actually, the gap equations (\ref{eq:Deeq}), (\ref{eq:la}),  and (\ref{eq:V0})
were obtained in Refs.~\cite{LS,LMS} in limit of weak interlayer interactions ($r_s\ll 1$ or $p_F d \ll 1$)
within the RPA approximation before Eq.~(\ref{eq:Tc}) for the mean-field temperature was derived in Ref.~\cite{KE} using the large-$N$ approximation.
An important conclusion of Ref.~\cite{KE} is that due to the reasonably large numerical value $N=8$, these equations
are valid and provide reliable quantitative prediction for $T_c$ not only for weak interlayer interactions, but also
in the realistic case of moderate to strong interactions, when $r_s \gtrsim 1$.

Let us estimate the accuracy of the result (\ref{eq:Tcmax2}).
This equation is obtained by taking into account the leading in $N$
diagrams (Fig.~\ref{fig:largeN}) for the effective interlayer interaction.
The diagrams that are neglected [such as diagrams (3.2) and (3.3) compared to (3.1)] are smaller in $1/N$ than the
leading contribution. Therefore, one could represent the exact
coupling constant $\lambda _{\Delta =0}^{{\rm max}}$ in
the form of an asymptotic series in $N$ with Eq.~(\ref{eq:lamax})
giving the leading term of the series. This way,
one can write down the inverse coupling constant as
\begin{equation}
1/\lambda _{\Delta =0}^{{\rm max}}=2N-\mathcal{O}(1), \label{eq:lamaxpr}
\end{equation}%
where $\mathcal{O}(1)>0$ represents all subleading in $1/N$ terms and is
of order unity, $\mathcal{O}(1)\sim 1$.

Inserting Eq.~(\ref{eq:lamaxpr}) into the exponential function in
Eq.~(\ref{eq:Tc})  one can conclude that
in the best case scenario the transition temperature
$T_{c}^{{\rm max}}$ could be about one order ($e^{\mathcal{O}(1)}$) greater than
the one  given by Eq.~(\ref{eq:Tcmax2}). With the proper (yet unknown within the logarithmic accuracy)
numerical prefactor in Eq.~(\ref{eq:Tcmax2}) one could optimistically
hope that $T_{c}^{{\rm max}}$ could reach $100{\rm mK}$ for $\epsilon_F \sim 0.3 {\rm eV}$.
Unfortunately, what concerns the possibility to observe the
effect experimentally, this would not radically change the situation.
First, 
as it was mentioned, the maximum (\ref{eq:lamax}) of the interaction strength
is virtually unachievable in the experiment, since this requires the
condition (\ref{eq:cond}) to be satisfied. Second, as it will be discussed in
Sec.~\ref{sec:disorder}, inevitable boundary scattering at the edges of graphene sheets will destroy any condensate
with $T_{c}\lesssim 1{\rm K}$ even in a system free of any bulk disorder.

\section{Excitonic gap and screening in the excitonic state \label{sec:BKT}}

\begin{figure}
\includegraphics[width=.50\textwidth]{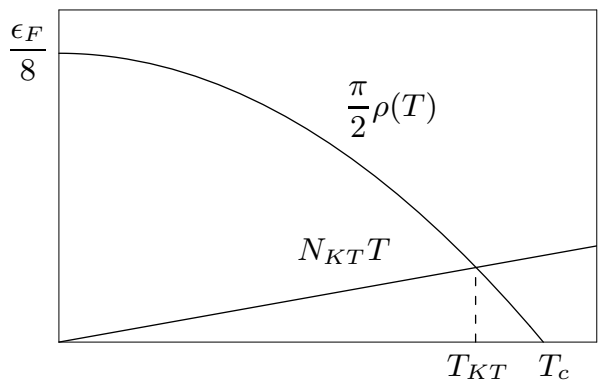}
\caption{ Graphical solution of Eq.~(\ref{eq:TBKTeq}) for the BKT transition temperature $T_{KT}$. The stiffness $\rho(T)$ [Eq.~(\ref{eq:rho})]
of the order parameter takes a universal value $\rho(T=0)=\epsilon_F/(4 \pi)$  at low temperatures (when $T \ll \De(T)$)
decreases with increasing  temperature $T$, and  vanishes at the mean-field transition temperature $T=T_c$, where
the excitonic gap $\De(T=T_c)=0$ turns to zero. Therefore, the BKT temperature $T_{KT}$ cannot exceed the mean-field temperature $T_c$, Eq.~(\ref{eq:TBKTTc}). } \label{fig:KT}
\end{figure}

In addition to the mean field temperature $T_{c}$ obtained in the previous section, one can
estimate the Berezinski-Kosterlitz-Thouless (BKT) transition temperature.
The BKT transition~\cite{Bereza,KT} occurs in two-dimensional superfluid systems,
in which thermal fluctuations destroy the long-range order of the order parameter.
The BKT temperature is defined as the temperature at which creation of
topological defects (most commonly, vortices) in the order parameter becomes thermodynamically favorable.
Mathematically, this condition can  be expressed as:
\beq
    \frac{\pi}{2}\rho(T_{KT})= N_{KT} T_{KT},
\label{eq:TBKTeq}
\eeq
where $\rho(T)$ is the so-called stiffness of the order parameter, which determines the energy of the topological defect.
The factor $N_{KT}=1$ for vortices. As shown in Ref.~\cite{AKT}, however, in graphene  creation of halfvortices
can be more favorable, in which case $N_{KT}=4$ and $T_{KT}$ would be somewhat lower than for $N_{KT}=1$.
Within the mean-field theory, the stiffness has the  form~\cite{AKT}
\beq
    \rho(T)= \frac{\epsilon_F}{4}  T \sum_{n=-\infty}^{+\infty} \frac{\De^2(T)}{[\eps^2_n+\De^2(T)]^{3/2}},
\label{eq:rho}
\eeq
where $\De(T)$ is the excitonic gap, which should be obtained by solving the gap equation (\ref{eq:Deeq}).
The stiffness (\ref{eq:rho}) takes a universal value $\rho(T=0)=\epsilon_F/(4 \pi)$ at low temperatures, when $T \ll \De(T)$,
decreases with increasing temperature, and vanishes when $T$ approaches the mean-field  temperature (\ref{eq:Tc}),
$T \rtarr T_c$, where the gap turns to zero, $\De(T_c)=0$. Therefore, the BKT temperature $T_{KT}$, which is a solution of Eq.~(\ref{eq:TBKTeq}), cannot exceed the mean-field temperature $T_c$,
\beq
    T_{KT} < T_c.
\label{eq:TBKTTc}
\eeq
This is illustrated graphically in Fig.~\ref{fig:KT}.
The result (\ref{eq:TBKTTc})
is also intuitively clear,
since one would naturally expect thermal fluctuations of the order parameter to suppress, rather than enhance,
the superfluidity of the system.

One can conclude from Eq.~(\ref{eq:TBKTTc}) that the estimate (\ref{eq:Tcmax}), obtained within the mean-field approach,
sets the upper bound for the temperature of any realistic~\cite{AKT} phase transition
into the excitonic state in a double-layer graphene system.
Nevertheless, as mentioned in the introduction, the authors of Ref.~\cite{MDc} came to quite a different conclusion.
On the one hand, they agreed~\footnote{
The authors of Ref.~\cite{MDc} actually provided an estimate
$T_{c}\sim 6 \cdot 10^{-5} \epsilon_F $ for the mean-field temperature,
which is
more than two orders greater the result (\ref{eq:Tcmax}) for the highest possible value of $T_c$. It is hard to comment on
the possible origin of this discrepancy, since no analytical formula for $T_{c}$ was presented in Ref.~\cite{MDc}.
} that the mean-field
temperature $T_c$ indeed appears to be significantly smaller than their initial estimate $
T_{KT}^{\ast }$~[Eq.~(\ref{eq:TBKT*})] for the BKT temperature, provided  screening is
taken into account.
 On the other hand,
they argued 
that in the excitonic phase the gap in the excitation spectrum
suppresses screening, which would mean that the gap $\De^*(T)$ itself
should be of the same order as the estimate (\ref{eq:TBKT*}) obtained neglecting screening,
see Eq.~(\ref{eq:De*}).
Such a strong discrepancy between $T_c$~[Eq.~(\ref{eq:Tcmax})] and $\De^*(T=0)$~[Eq.~(\ref{eq:De*})] indicates, according to Ref.~\cite{MDc},
that the system should
undergo a first-order transition into the excitonic state at temperature $T \sim \De^*(T=0)$, which would again correspond to
high temperatures.

We will now  show that, in fact, the other, less exciting,  scenario is realized.
We will demonstrate that the suppression of screening by the gap in the excitonic state,
although it does take place, has a negligible effect on the value
of the gap, when the latter is determined
self-consistently from the gap equations (\ref{eq:Deeq}) and (\ref{eq:la}).
As a result, the system  is in the weak-coupling regime of  the screened interactions,  described by the large-$N$  approximation, at all temperatures below $T_c$ and the value of the gap $\De(T)\lesssim 10^{-7} \epsilon_F$ in the excitonic state appears to be in full
agreement with the value (\ref{eq:Tcmax}) of the mean-field temperature.

To prove this point, one needs the expression for the screened
interlayer Coulomb potential $V_{\Delta }(q)$ in the excitonic
phase, where  $\Delta \neq 0$. Performing the calculations analogous to those of RPA, in the large-$N$ approximation
one obtains
\begin{equation}
V_{\Delta }(q)=\frac{V_{+}(q)}{1+2\Pi _{\Delta }^{+}(q)V_{+}(q)}-\frac{%
V_{-}(q)}{1+2\Pi _{\Delta }^{-}(q)V_{-}(q)}.  \label{eq:VDe}
\end{equation}%
Here
\begin{equation}
V_{\pm }(q)=\frac{1}{2}[1\pm \exp (-qd)]\frac{2\pi e_{\ast }^{2}}{q}
\label{eq:Vpm}
\end{equation}%
are the half-sum and half-difference of the intralayer $V_{0}(q)$ and
interlayer $V_{1}(q)$ bare Coulomb potentials~[Eq.~(\ref{eq:Vc})]. The
polarization operators
\begin{equation}
\Pi _{\Delta }^{\pm }(q)=\Pi _{\Delta }^{0}(q)\pm \Pi _{\Delta }^{1}(q)
\end{equation}%
in Eq.~(\ref{eq:VDe}) are the sum and difference of the intralayer $\Pi
_{\Delta }^{0}(q)$ and interlayer $\Pi _{\Delta }^{1}(q)$ polarization
operators. The operators $\Pi _{\Delta }^{0}(q)$ and $\Pi _{\Delta }^{1}(q)$
describe 
the density response in one layer to the density perturbation in the same
and other layer, respectively, and can be
written down in the Matsubara representation~\cite{AGD} as
\beq
    \Pi_\De^{0,1}(q)= \int_0^{1/T} \dt \tau \int \dt \rb \, e^{-{\rm i} \qb \rb} \lan \rhoh_i (\tau,\rb) \rhoh_j(0,0) \ran
\label{eq:Pidef}
\eeq
with the same layer indices ($i=j$) for $\Pi_\De^{0}(q)$ and different ones ($i\neq j$) for $\Pi_\De^{1}(q)$.
In  Eq.~(\ref{eq:Pidef}), $\rhoh_i (\rb)=\phih_i^\dg(\rb) \phih_i(\rb)$  are the density operators and  $\tau$ is the Matsubara time.

The polarization operators (\ref{eq:Pidef}) can be calculated using the Green's functions for the excitonic state.
Doing so, one obtains that in the excitonic state  $\Pi _{\Delta
}^{-}(q)$ does not depend on the gap at all and is equal to its normal state
value
\begin{equation}
\Pi _{\Delta }^{-}(q)=N_{s}N_{v}\nu .
\end{equation}%
On the hand, the operator $\Pi _{\Delta }^{+}(q)$ is affected by the gap and the
expression for it reads
\beq
    \Pi_\De^+(q)=N_s N_v \nu \lt\{1-\pi T \sum_{n=-\infty}^{+\infty} \int_{0}^{2\pi} \frac{{\rm d} \theta}{2\pi} \frac{\De^2}{[(\eps_n^2+\De^2)+(v q \cos \theta)^2/4] \sqrt{\eps^2+\De^2}} \rt\}
\label{eq:PiDe}
\eeq
Here, as in Eq.~(\ref{eq:Deeq}), $\varepsilon _{n}=\pi T(2n+1)$
are the fermionic Matsubara frequencies and the summation is done over all
integer $n$. 

Aiming to find the highest possible value of the gap, let us again consider the
limit (\ref{eq:cond}) of \textquotedblleft zero\textquotedblright\
interlayer distance, when the potential $V_{\Delta }(q)$~[Eq.~(\ref{eq:VDe}%
)] is maximal. In this limit the second term in Eq.~(\ref{eq:VDe})
can be neglected, $V_{+}(q)=V_{0}(q)$ [Eqs.~(\ref{eq:Vc}) and (\ref{eq:Vpm}%
)], and the interlayer interaction potential takes the form
\begin{equation}
V_{\Delta }^{{\rm max}}(q)=\frac{V_{0}(q)}{1+2\Pi _{\Delta }^{+}(q)V_{0}(q)}%
.  \label{eq:VDemax}
\end{equation}%
Again, as in Eq.~(\ref{eq:V0max}), the factor $2$ in the denominator
in Eq.~(\ref{eq:VDemax}) originates from summing the polarization
operators of both layers.

Let us discuss the properties of the polarization operator (\ref{eq:PiDe})
and the corresponding properties of the potential (\ref{eq:VDemax}). In the
normal state ($\De=0$), the second term in Eq.~(\ref{eq:PiDe}) is zero, $\Pi _{\Delta
=0}^{+}(q)$ is given by Eq.~(\ref{eq:Pi0}) and one recovers the expression (%
\ref{eq:V0max}) for $V_{\Delta=0 }^{{\rm max}}(q)$ from Eq.~(\ref{eq:VDemax}%
).
The second term in Eq.~(\ref{eq:PiDe}) exists in the excitonic
phase only, when $\Delta \neq 0$. It is a decreasing function of $q$ and
approaches zero  for $q\gg \Delta /v$ exceeding the condensate momentum $\De/v$. So,
for momenta $q\gg\Delta /v$ the polarization operator
\begin{equation}
\Pi _{\Delta }^{+}(q\gg \Delta /v)\approx N_{s}N_{v}\nu
\label{eq:PiDen}
\end{equation}%
is given by its normal state value (\ref{eq:Pi0}) at any temperature, including $T=0$, and
the interlayer interaction potential is completely screened,
\begin{equation}
\nu V_{\Delta }^{{\rm max}}(q\gg \Delta /v)\approx \frac{1}{N}.
\label{eq:VDemaxn}
\end{equation}%
Therefore, an important conclusion that can be drawn from Eqs.~(\ref{eq:PiDen}) and (\ref{eq:VDemaxn}) is that for momenta  $ q $
exceeding the condensate momentum $\De /v $ screening is just as strong in the excitonic state as it is in the normal state.

\begin{figure}
\includegraphics[width=.50\textwidth]{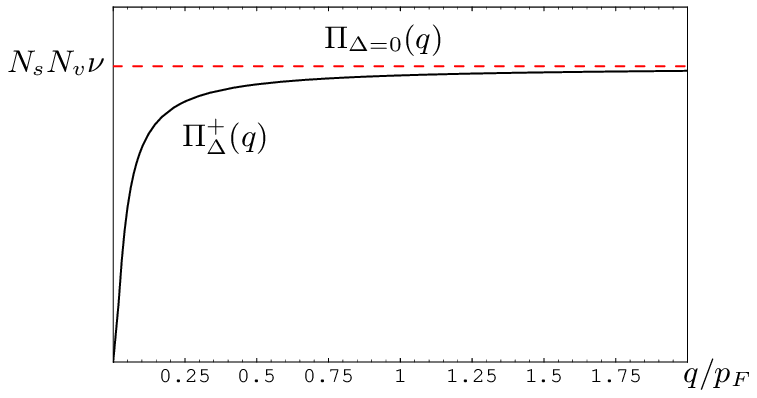}
\includegraphics[width=.475\textwidth]{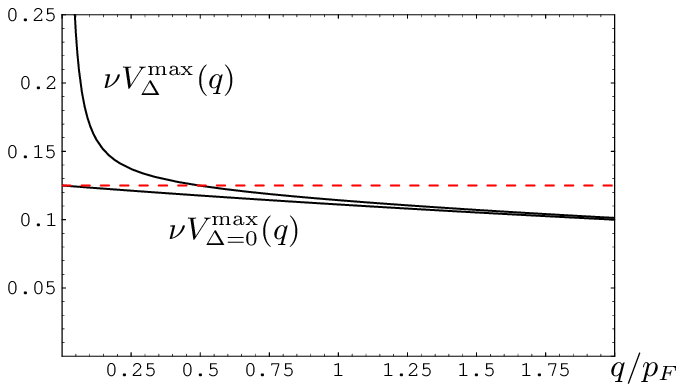}
\caption{ Dependence of the polarization operator  $\Pi_{\De}^+(q)$~[Eq.~(\ref{eq:PiDe}), left graph] and the maximum of the screened interlayer potential $V_{\De}^{\rm max}(q)$~[Eq.~(\ref{eq:VDemax}), right graph] on momentum $q$  in the excitonic state. The case of $\De/\epsilon_F=10^{-2}$, $r_s=1$, and zero temperature is shown. The operator $\Pi_{\De}^+(q)$ is suppressed  for $q \lesssim \De/v$
below the condensate momentum $\De/v$, whereas for greater $q \gg \De/v$ it
approaches the normal-state value $\Pi_{\De=0}(q)=\nu N_s N_v$~[Eq.~(\ref{eq:Pi0}), dashed line].
Consequently, the potential $\nu V_{\De}^{\rm max}(q)$ is unscreened and exceeds its small normal-state value $\nu V_{\De=0}^{\rm max}(q) \leq 1/8$~[Eqs.~(\ref{eq:V0max}) and (\ref{eq:V0maxq0})] only for $q \lesssim \De/v$.
As the excitonic gap $\De$ is determined by all momenta in the range $0 \leq q \leq 2 p_F$, see Eqs.~(\ref{eq:Deeq}) and (\ref{eq:la}),
the suppression of screening at $q \lesssim \De/v$ gives only a minor correction to the coupling strength $\la_\De$,
which is small in the ratio $\De/\epsilon_F$~[Eq.~(\ref{eq:dela})]. }
\label{fig:PiDe}
\end{figure}

At the same time, the situation is different for small momenta. The
polarization operator $\Pi _{\Delta }^{\pm }(q)$ is indeed suppressed at
$q\lesssim \Delta /v$, when the second term in Eq.~(\ref%
{eq:PiDe}) contributes considerably. The smaller $q$ and $T$ are, the
stronger the suppression of $\Pi _{\Delta }^{+}(q)$ is.
In fact, at zero temperature $T=0$ and zero momentum $q=0$ the second
term in Eq.~(\ref{eq:PiDe}) cancels the first term exactly and one
gets
\begin{equation}
\Pi _{\Delta }^{+}(q=0)|_{T/\Delta =0}=0.  \label{eq:PiDeq0}
\end{equation}%
This means that at small momenta and low temperature screening is indeed suppressed and  the interlayer Coulomb potential is unscreened,
\begin{equation}
 V_{\Delta }^{{\rm max}}(q\lesssim \Delta /v)|_{T\ll \Delta} \approx \frac{2 \pi e_*^2}{q}.
\label{eq:VDemaxq0}
\end{equation}%
Equation~(\ref{eq:PiDeq0})
has a simple physical
interpretation: at zero momentum and temperature all electrons and holes are paired
into dipoles and there are no free carriers (excitations) that could screen the
interaction.

The dependence of the polarization operator $\Pi _{\Delta }^{+}(q)$~[Eq.~(%
\ref{eq:PiDe})] and of the interaction potential $V_{\Delta }^{{\rm max}%
}(q)$~[Eq.~(\ref{eq:VDemax})] on $q$, illustrating the above properties, is plotted in Fig.~%
\ref{fig:PiDe}.

It is essentially the property (\ref{eq:PiDeq0}) of the polarization operator (\ref{eq:PiDe})
that the authors of Ref.~%
\cite{MD} used in Ref.~\cite{MDc} to argue that, in terms of its effect on
the value of the gap $\Delta $, screening should be suppressed. The
expression (\ref{eq:PiDe}) at $q=0$ 
can be rewritten in the form of Eq.~(4) of Ref.~\cite{MDc}
using the conventional method~\cite{AGD} of replacing the sum over the
Matsubara frequencies $\varepsilon _{n}$ by the integral over a continuous
energy variable. %
According to Ref.~\cite{MDc}, the property (\ref{eq:PiDeq0}) justifies the
neglect of screening,
when
determining the value of the gap in the condensed phase.

It is easy to understand now from Eqs.~(\ref{eq:Deeq}), (\ref{eq:la}), (\ref%
{eq:PiDe}) and (\ref{eq:VDemax}), and the discussed properties
of the polarization operator (\ref{eq:PiDe}) that the suppression of
screening at small momenta $q\lesssim \Delta /v$ has a very little,
in fact, negligible effect on the value of the gap. Indeed, as follows from
Eq.~(\ref{eq:la}) and was pointed out earlier in Sec.~\ref{sec:gap}, the
coupling strength $\lambda _{\Delta }$ and, consequently, the gap $\Delta $%
~[Eq.~(\ref{eq:Deeq})], is determined by all scattering momenta $%
q=2p_{F}\sin \theta /2$ in the range $0\leq q\leq 2p_{F}$.
As we have just discussed, screening is suppressed by
the gap in the range $0\leq q\lesssim \Delta /v$ only~[Eqs.~(\ref{eq:PiDeq0}) and (\ref{eq:VDemaxq0})], whereas in the range $%
\Delta /v\lesssim q\leq 2p_{F}$ screening is just as strong as in the normal
state~[Eqs.~(\ref{eq:PiDen}) and (\ref{eq:VDemaxn}), Fig.~\ref{fig:PiDe}]. Therefore, the
difference
\[\delta \lambda _{\Delta }=\lambda _{\Delta }-\lambda _{\Delta =0}
\] between the pairing constant $\lambda _{\Delta }$~[Eq.~(\ref{eq:la})]
in the condensed phase
and its value $\lambda _{\Delta =0}$ in the normal state comes from the range of
momenta $0\leq q\lesssim \Delta /v$ only
and 
is small in the ratio of the gap $\Delta $ to the Fermi energy $\epsilon
_{F}=vp_{F}$,
\begin{equation}
\delta \lambda _{\Delta }/\lambda _{\Delta =0}\sim \Delta /\epsilon _{F}.
\label{eq:dela}
\end{equation}%

Equation~(\ref{eq:dela}) shows unambiguously that the presence of the
excitonic gap does  not influence the value of the coupling constant as long as $\De \ll \epsilon_F$. In other
words, the statements of  Ref. \cite{MDc} cannot be correct unless the
excitonic gap becomes of the order of the Fermi energy.

Let us now find the maximum value $\De_0^{\rm max}$ [in the limit (\ref{eq:cond}) of ``zero'' interlayer distance]
of the gap $\Delta _{0}=\Delta
(T=0)$ at zero temperature, when the suppression of screening by the gap is greatest.
From Eq.~(\ref{eq:Deeq}), one obtains that $%
\Delta _{0}$ satisfies the equation
\begin{equation}
\Delta _{0}\approx \exp (-1/\lambda _{\Delta _{0}})\,\epsilon _{F}.
\label{eq:De0eq}
\end{equation}%
Unlike Eq.~(\ref{eq:Tc}), which provides the explicit expression
for the mean field transition temperature, one has to solve Eq.~(%
\ref{eq:De0eq}) to obtain $\Delta _{0}$, since $\lambda _{\Delta _{0}}$
depends on $\Delta _{0}$.
However, considering the extremely small
numerical value of $T_{c}^{{\rm max}}/\epsilon _{F}$~[Eq.~(\ref{eq:Tcmax2}%
)]
we conclude immediately that
\begin{equation}
\Delta _{0}^{{\rm max}}\approx \exp (-1/\lambda _{\Delta =0}^{{\rm max}%
})\,\epsilon _{F}=\exp (-16)\,\epsilon _{F}\approx 10^{-7}\,\epsilon _{F},
\label{eq:De0max}
\end{equation}%
for which the suppression of screening by the gap is neglected
completely, is the solution of Eq.~(\ref{eq:De0eq}). Indeed, inserting the value (\ref{eq:De0max})
into Eq.~(\ref{eq:la}) and calculating the integrals numerically, we obtain that $\delta \lambda _{\Delta _{0}}^{%
{\rm max}}/\lambda _{\Delta =0} \sim 10^{-6}$ [Eq.~(\ref{eq:dela})], and
therefore $\De_0^{\rm max}$ is determined by Eq.~(\ref{eq:De0max}) with the relative precision $\sim 2 N \delta \lambda _{\Delta _{0}}^{%
{\rm max}}/\lambda _{\Delta =0} \sim 10^{-4}$.

Besides Eq.~(\ref{eq:De0max}), there are  no other solutions to Eq.~(\ref{eq:De0eq})
in the range $\Delta_0 \lesssim \epsilon _{F}$, which can be checked numerically.
One could use the highest hypothetical  value  (\ref{eq:De*}) of the gap as the zeroth approximation, insert it into the right-hand side of Eq.~(\ref{eq:De0eq}), and obtain that already on the second iteration
the solution collapses to the value (\ref{eq:De0max}) with high precision.

We conclude that, contrary of the arguments of Ref.~\cite{MDc}, in terms of
its effect on the excitonic pairing, screening is as strong in
the excitonic state as it is in the normal one.
Partial suppression of screening at small momenta $q\lesssim
\Delta /v$ has a negligible effect on the value of the gap when the latter
is determined self-consistently from the gap equation. This clearly
follows from the fact that all the scattering momenta $q$ in the
range $q\leq 2p_{F}$ contribute to the value of the gap, and not only small
ones. Therefore, the normal-state expression~(\ref{eq:V0}) for the screened intrelayer Coulomb potential
can be used with great precision to study the excitonic phase at any temperature $T \leq T_c$
below the mean-transition temperature $T_c$ (\ref{eq:Tcmax}).
As a result, 
the maximum possible value of the zero-temperature gap (\ref{eq:De0max})
appears to be in full agreement with the result (\ref{eq:Tcmax}) for the
mean field transition temperature. The fact  that $\exp (-16)\epsilon _{F}\sim
10^{-7}\epsilon _{F}$ is the only energy scale of the excitonic
pairing in the double-layer graphene system
proves that Eq.~(\ref{eq:Tcmax}) really sets the upper bound for the transition temperature
and invalidates the conjecture of Ref.~\cite{MDc} about the high-temperature first order transition.

We would like to stress that all the above discussion about how
screening is affected by the gap in the excitonic state was
not actually in any way specific to graphene. The very same analysis could
have been carried out for conventional systems~\cite{KK,LY} with quadratic
electron spectrum.
What concerns graphene, the point that
screening is only negligibly affected by the gap has been made
earlier in Ref.~\cite{AKT}, 
where the dependence of the polarization operator on $\Delta $ in the
excitonic state (its suppression in the minor range of momenta $%
q\lesssim \Delta /v$) was neglected in the calculations for the reasons
discussed above. 

\section{Influence of disorder \label{sec:disorder}}

\begin{figure}[tbp]
\includegraphics[width=.50\textwidth]{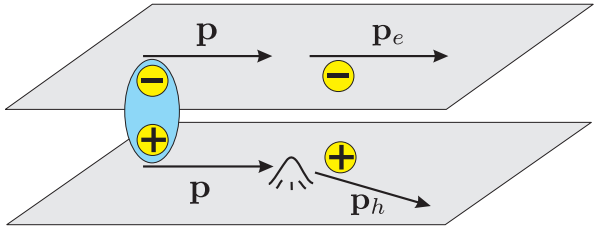}
\caption{ Sensitivity of the excitonic condensate to the impurity
scattering. Impurities with the size of potential smaller than the
interlayer distance scatter electrons and holes not identically,
thereby breaking electron-hole pairs and suppressing the
condensate.} \label{fig:scattering}
\end{figure}

Although the estimate (\ref{eq:Tcmax})
predicts a really small value of the transition temperature,
one could still hope to observe excitonic effects in a $1{\rm mK}$ range.
Unfortunately, in a realistic system even at these low temperatures the excitonic pairing could
hardly be realized. The reason for this is that these
estimates were made for a system that is free of any kind of
disorder, while 
the excitonic condensate in double-layer systems is sensitive to the impurity scattering~\cite%
{Zittartz,LY}. Indeed, since the bound electron and hole carry the same
momentum $\mathbf{p}$, any scattering process that changes the momentum of
electron and hole not identically, i.e., $\mathbf{p}\rightarrow \mathbf{p}%
_{e}$ for electron and $\mathbf{p}\rightarrow \mathbf{p}_{h}$ for hole, so
that $\mathbf{p}_{e}\neq \mathbf{p}_{h}$, breaks the electron-hole pair (see
Fig.~\ref{fig:scattering}).
This is the case for
impurities with the range of the scattering potential less than
the interlayer distance $d$, since the potential of such
impurities differs in the two layers. Therefore, any sufficiently strong-range impurities destroy the excitonic condensate.

Originally, the effect of the impurity scattering on the excitonic
condensate was studied analytically for conventional systems in Refs.~\cite%
{Zittartz,LY}. The theory is formally analogous to the Abrikosov-Gorkov's
theory~\cite{AG} for magnetic impurities in superconductors. For a double
layer graphene system, analogous calculations were performed in Ref.~\cite%
{MDdisorder}. The main result is that sufficiently short-range impurities
with the scattering time $\tau $ destroy the excitonic condensate completely
as soon as
\begin{equation}
\tau T_{c} \lesssim 1 ,  \label{eq:disorder}
\end{equation}%
where $T_{c}$ is the transition temperature of the ideally clean system.
Equivalently, for the condensate to exist, electron momentum has to be
conserved at the scale of the correlation length $ v/T_{c}$. How weak
impurity scattering in a realistic system can be? Even if the system is free
of bulk disorder, the mean free path is limited by the sample size due to
the boundary scattering at the sample edges.
With the typical size of graphene devices $\sim 1\mu {\rm m}$ this
corresponds to the scattering rate $1 /\tau \sim 1{\rm K}$. Therefore,
we can conclude that any condensate with $T_{c}$ in a clean system below $1%
{\rm K}$ would be completely destroyed in a realistic system.

\section{Comparison with conventional double-layer systems}

In terms of engineering properties, the double-layer graphene system could seem as an
exceptional candidate for the realization of the excitonic condensation.
However, as we have seen in the previous sections, the peculiar properties
of graphene spectrum (additional valley degrees of freedom and chirality)
appear to play a negative role for this effect.
To illustrate this point,
let us estimate for comparison
the highest possible value of the transition temperature in a double-layer system
with the conventional quadratic electron spectrum, such as a semiconductor double quantum well, using the large-$N$
approximation.

How should the equations of Sec.~\ref{sec:gap} be modified in this case?
First, since the valley degeneracy $N_{v}=2$ is absent, the total number of flavors is $%
N=N_{s}N_{l}=4$ due to the two ($N_{s}=2$) projections of spin and two
layers ($N_{l}=2$), Therefore, the maximum of the dimensionless coupling
constant equals
\begin{equation}
\nu V_{\Delta =0}^{{\rm max}}(q=0)=\frac{1}{N}=\frac{1}{4}
\label{eq:V0maxconv}
\end{equation}%
%
Second, the factor $(1+\cos \theta )/2$, which stems from the chiral nature
of Dirac quasiparticles and results in the suppression of backscattering, is
absent in Eq.~(\ref{eq:la}).
Therefore, the maximum coupling strength $\lambda _{\Delta =0}^{{\rm max}}$
that determines the gap according to Eq.~(\ref{eq:Deeq}) is just equal to Eq.~(\ref{eq:V0maxconv}),
\begin{equation}
\lambda _{\Delta =0}^{{\rm max}}=\frac{1}{N}=\frac{1}{4}.
\end{equation}%
As we see, the pairing strength $\lambda _{\Delta =0}^{{\rm max}}$
in a double-layer with conventional spectrum is four times greater
than the one in graphene-based system~[Eq.~(\ref{eq:lamax})]. According to Eq.~(\ref{eq:Tc}), for the maximum of
the transition temperature one obtains
\begin{equation}
T_{c}^{{\rm max}}\approx \exp (-N)\epsilon _{F}=\exp (-4)\epsilon
_{F}\approx 10^{-2}\epsilon _{F}.  \label{eq:Tcmaxconv}
\end{equation}%
This value is five orders higher than that for the graphene double-layer~ [Eq.~(\ref{eq:Tcmax})]
and by no means does it lead to similar pessimistic conclusions about the feasibility of the effect. Of
course, in this case  the estimate (\ref{eq:Tcmaxconv}) may already be quite crude
because $\lambda _{\Delta =0}^{{\rm max}}=1/4$ is not that small. However, this only indicates
and the system could be on the verge of applicability of the weak-coupling limit and the actual $T_c$ could be even higher than
predicted by Eq.~(\ref{eq:Tcmaxconv}).

\section{Excitonic condensation in a single-layer graphene}

Besides a double-layer graphene system (Fig.~\ref{fig:system}),
there were a couple of theoretical proposals how excitonic pairing could be realized in a single layer of graphene.
We briefly discuss  them in this section.

One possibility~\cite{AKT} would be to apply a strong in-plane magnetic field to a single layer of
graphene. Such a field acts on the spins of
the carriers only, and the Zeeman splitting creates electrons with
one spin polarization and holes with the opposite polarization in
an initially neutral sample.
As electrons and holes in such a system
have well-defined Fermi surfaces, the transition temperature can be estimated within the large-$N$ approximation using
Eqs.~(\ref{eq:Deeq}) and (\ref{eq:la}) of the BCS-like theory.
The number of species in a single-layer graphene is $N_1=N_s N_v=4$ and
the maximum of the coupling constant (\ref{eq:la}) equals to
\beq
    \lambda_{\De=0}^{\rm max}=\frac{1}{2\,N_{s}\,N_{v}}=\frac{1}{8}
\eeq
The Zeeman splitting energy $\epsilon_Z$ plays the role of the Fermi energy in such a setup.
Therefore, for the maximum of the transition temperature one obtains
\beq
    T_c^{\rm max} \approx  \exp (-\lambda _{\De=0}^{{\rm max}})\, \epsilon_Z=\exp (-8)\, \epsilon_Z \approx 3\cdot 10^{-4} \epsilon_Z
\label{eq:Tcmax1}
\eeq
The exponential factor $\exp(-8)$ is not as small as  $\exp(-16)$ in the double-layer system. However, the Zeeman splitting
energy  cannot be extremely high even for experimentally very high
magnetic fields $B$. For $B\approx 40\,{\rm T}$ one can estimate from Eq.~(\ref{eq:Tcmax1})
$T_{c}^{{\rm max}}\sim 20 {\rm mK}$. Alternatively, instead of applying external magnetic field, one could bring
the graphene sheet in contact with the ferromagnet, in which case the Zeeman splitting would be caused by the ferromagnet
exchange field~\cite{AKT}.

In the electrically biased double-layer graphene and a single-layer graphene in the parallel magnetic field electrons and holes
have a finite density of states at the mutual Fermi level. This way the excitonic pairing is realized
though the Cooper instability mechanism, and a finite transition temperature  formally exists for arbitrary small strength of the bare Coulomb interactions $r_s$
(although, as it was demonstrated here, even for moderate to strong interactions the transition temperature appears to be numerically extremely small).
It was conjectured in a number of works~\cite{Khvesh,GGMS,Khvesh2,Khvesh3} that the requirement of the finite DoS in actually not necessary and that excitonic pairing could be  possible in an undoped graphene layer, with the chemical potential
at the Dirac point.
The analysis of  Refs.~\cite{Khvesh,GGMS,Khvesh2,Khvesh3} predicts that opening of the excitonic gap at the Dirac point may occur,
provided the Coulomb coupling strength $r_s=e^2/(\eps v)$ exceeds certain threshold value $r_s^*$. This value depends on the
number of flavors and for $N=N_1=4$ equals  $r_s^*=2.33$,
which is quite close to $r_s \approx 2.16$ in vacuum ($\eps=1$).

\section{Conclusion}

We theoretically studied the possibility of excitonic condensation in a recently
proposed double-layer graphene system.
On the one hand, this system possesses several obvious engineering
advantages compared to semiconductor double quantum wells, used
so far experimentally. On the other
hand, 
the properties of the graphene spectrum
appear to be very unfavorable for realization of excitonic
condensation in such a  system.
Additional valley degrees of freedom and suppression of backscattering in graphene
lead to a much smaller value  of the screened Coulomb interaction strength than in
a system with conventional electron spectrum.
It is shown the effect of screening on the excitonic pairing
is just as strong in the excitonic state as it is in the normal state.
For this reason, the value of the excitonic gap  is found to fully agree
with the previously obtained~\cite{KE} estimate  for the mean-field transition temperature, the maximum
possible value $\sim 10^{-7} \epsilon_F$ of both being in the $1{\rm mK}$ range for a perfectly clean system.
These findings disprove the predictions~\cite{MD,ZJ,MDc}  of the excitonic condensation at room temperatures and,  unfortunately, render experimental observation of this undoubtedly interesting effect in a double-layer graphene system unlikely.

\section*{Acknowledgements}

Insightful discussions with Igor Aleiner are greatly appreciated.


\section*{References}
\begin{thebibliography}{99}
\bibitem{KK} L. V. Keldysh and Yu. V. Kopaev, Sov. Phys. Solid State \textbf{%
6}, 2219 (1965).





\bibitem{LY} Yu. E. Lozovik and V. I. Yudson, JETP Lett. \textbf{22}, 274
(1975); Sov. Phys. JETP \textbf{44}, 389 (1975); Solid State Commun. \textbf{%
19}, 391 (1976); Solid State Commun. \textbf{21}, 211 (1977).


\bibitem{exp1} U. Sivan, P. M. Solomon, and H. Shtrikman, Phys. Rev. Lett.
\textbf{68}, 1196 (1992).

\bibitem{exp11} A. F. Croxall, K. Das Gupta, C. A. Nicoll, M. Thangaraj, H. E. Beere, I. Farrer, D. A. Ritchie, and M. Pepper
Phys. Rev. Lett. {\bf 101}, 246801 (2008).

\bibitem{exp12} J. A. Seamons, C. P. Morath, J. L. Reno, and M. P. Lilly, Phys. Rev. Lett. {\bf  102}, 026804 (2009).

\bibitem{exp2} L. V. Butov, A. Zrenner, G. Abstreiter, G. B{\"o}hm, and G.
Weimann, Phys. Rev. Lett. \textbf{73}, 304 (1994).

\bibitem{exp3} M. Bayer, V. B. Timofeev, F. Faller, T. Gutbrod, and A.
Forchel, Phys. Rev. B \textbf{54}, 8799 (1996).

\bibitem{exp4} A. A. Dremin, V. B. Timofeev, A. V. Larionov, J. Hvam, and K. Soerensen,
JETP Lett. {\bf 76}, 450 (2002).

\bibitem{exp5} L. V. Butov, C. W. Lai, A. L. Ivanov, A. C. Gossard, and  D. S. Chemla, Nature {\bf 417}, 47 (2002).

\bibitem{exp6} L. V. Butov, A. C. Gossard, and  D. S. Chemla, Nature {\bf 418}, 751 (2002).

\bibitem{Eisen} I. B. Spielman, J. P. Eisenstein, L. N. Pfeiffer, and K. W.
West, Phys. Rev. Lett. \textbf{84}, 5808 (2000); Phys. Rev. Lett. \textbf{87}%
, 036803 (2001);

\bibitem{EMD} J. P. Eisenstein and A. H. MacDonald, Nature \textbf{432}, 691
(2004).



\bibitem{gr1}
K. S. Novoselov, A. K. Geim, S. V. Morozov, D. Jiang, Y. Zhang, S. V. Dubonos, I. V. Grigorieva, and A. A. Firsov, Science {\bf 306}, 666 (2004).

\bibitem{gr2} K.S. Novoselov,
 A.K. Geim, S.V. Morozov, D. Jiang,
M.I. Katsnelson, I.V. Grigorieva, S.V. Dubonos, and A.A. Firsov,
Nature \textbf{438}, 197 (2005).

\bibitem{gr3} Y. Zhang, J. P. Small, M. E. S. Amori, and P. Kim, Phys. Rev. Lett. {\bf 94}, 176803 (2005).
\bibitem{gr4} Y. Zhang, Y.-W. Tan, H. Stormer, and   P. Kim, Nature {\bf 438}, 201 (2005).

\bibitem{gr5}  C. Berger, Z. Song, T. Li, X. Li, A. Y. Ogbazghi, R. Feng, Z. Dai, A. N. Marchenkov, E. H. Conrad, P. N. First, and W. A. de Heer, J. Phys. Chem. B {\bf 108}, 19912 (2004).

\bibitem{gr6}  C Berger, Z. Song, X. Li, X. Wu, N. Brown, C. Naud, D. Mayou, T. Li, J. Hass, A. N. Marchenkov, E. H. Conrad, P. N. First, and W. A. de Heer Science {\bf 312}, 1191 (2006).

\bibitem{grrise} A. K. Geim and K. S. Novoselov, Nat. Mater. {\bf 6}, 183 (2007).

\bibitem{LS} Yu.E. Lozovik and A.A. Sokolik, JETP Lett. {\bf 87},  55 (2008).

\bibitem{LMS} Yu. E. Lozovik, S. P. Merkulova, and A. A. Sokolik, Usp. Fiz.
Nauk \textbf{178}, 757 (2008), in Russian.

\bibitem{MD} H. Min, R. Bistritzer, J.-J. Su, and A.H. MacDonald, Phys. Rev.
B \textbf{78}, 121401(R) (2008). 

\bibitem{ZJ} C.-H. Zhang and Y. N. Joglekar, Phys. Rev. B \textbf{77},
233405 (2008).



\bibitem{Bereza} V. L. Berezinskii, Sov. Phys. JETP {\bf 32}, 493 (1971).

\bibitem{KT} J. M. Kosterlitz and D. J. Thouless, J. Phys. C {\bf 5}, L124 (1972).


\bibitem{KE}
M. Yu. Kharitonov and K. B. Efetov, Phys. Rev. B \textbf{78}, 241401 (2008).


\bibitem{MDc} R. Bistritzer, H. Min, J. J. Su, and A.H. MacDonald,
arXiv:0810.0331, unpublished.


\bibitem{AKT} I. L. Aleiner, D. E. Kharzeev, and A. M. Tsvelik, Phys. Rev. B
\textbf{76}, 195415 (2007).



\bibitem{YS} J. Ye and S. Sachdev, Phys. Rev. Lett. {\bf 80}, 5409 (1998).


\bibitem{Son} D. T. Son, Phys. Rev. B 75, 235423 (2007).

\bibitem{FA} M. S. Foster and I. L. Aleiner, Phys. Rev. B \textbf{77},
195413 (2008).

\bibitem{MFS}
M. Mueller, L. Fritz, and S. Sachdev, Phys. Rev. B {\bf 78}, 115406 (2008).



\bibitem{Khvesh} D. V. Khveshchenko, Phys. Rev. Lett. \textbf{87}, 206401
(2001);
{\em ibid} \textbf{87}, 246802 (2001).

\bibitem{GGMS} E. V. Gorbar, V. P. Gusynin, V. A. Miransky, and I. A.
Shovkovy, Phys. Rev. B \textbf{66}, 045108 (2002); Phys. Lett. A {\bf 313}, 472 (2003).

\bibitem{Khvesh2} D. V. Khveshchenko and H. Leal, Nucl. Phys. B {\bf 687}, 323 (2004);
D. V. Khveshchenko and W. F. Shively,
Phys. Rev. B {\bf 73}, 115104 (2006).

\bibitem{Khvesh3} D. V. Khveshchenko, J. Phys.: Condens. Matter {\bf 21}, 075303 (2009).


\bibitem{PO} B. Wunsch, T. Stauber, F. Sols, and F. Guinea, New J. Phys.
\textbf{8}, 318 (2006).



\bibitem{AGD}
A. A. Abrikosov, L. P. Gor'kov, and I. E. Dzyaloshinski, {\em ``Methods of Quantum Field Theory in Statistical Physics},
Prentice-Hall (1963).






\bibitem{Zittartz} J. Zittartz, Phys. Rev. \textbf{164}, 575 (1967).

\bibitem{AG} A. A. Abrikosov and L. P. Gor'kov, Sov. Phys. JETP {\bf 12}, 1243 (1961).

\bibitem{MDdisorder} R. Bistritzer and A.H. MacDonald,
Phys. Rev. Lett. {\bf 101}, 256406 (2008).





\end{thebibliography}
\end{document}